\documentclass[preprint]{aastex}
\usepackage{emulateapj5}
\usepackage{apjfonts}
\usepackage{natbib}

\input psfig.tex

\def\kms{km s$^{-1}$}

\def\asec{$^{\prime\prime}$}

\def\percm2{cm$^{-2}$}

\def\lax{{$\mathrel{\hbox{\rlap{\hbox{\lower4pt\hbox{$\sim$}}}\hbox{$<$}}}$}}
\def\gax{{$\mathrel{\hbox{\rlap{\hbox{\lower4pt\hbox{$\sim$}}}\hbox{$>$}}}$}}

\slugcomment{To appear in {\it The Astrophysical Journal Letters}}
\lefthead{Ravindranath, S. et al. }
\righthead{Evolution of Disk Galaxies in the GOODS-S Field}

\begin{document}

\title{Evolution of Disk Galaxies in the GOODS-South Field: Number Densities
and Size Distribution.\footnotemark[1]}

\footnotetext[1]{Based on observations obtained with the NASA/ESA 
{\it Hubble Space Telescope}, which is operated by the Association of 
Universities for Research in Astronomy, Inc. (AURA) under NASA contract 
NAS 5-26555 }

\author{
S. Ravindranath\altaffilmark{2}, H. C. Ferguson\altaffilmark{2,3},
C. Conselice\altaffilmark{4}, M. Giavalisco\altaffilmark{2},
M. Dickinson\altaffilmark{2,3}, E. Chatzichristou\altaffilmark{5},
D. de Mello\altaffilmark{6}, S. M. Fall\altaffilmark{2},
J. P. Gardner\altaffilmark{6}, N. A. Grogin\altaffilmark{3},
A. Hornschemeier\altaffilmark{3}, S. Jogee\altaffilmark{2},
A. Koekemoer\altaffilmark{2}, ~C. Kretchmer\altaffilmark{3},
M. Livio\altaffilmark{2}, B. Mobasher\altaffilmark{2},
R. Somerville\altaffilmark{2}\\
 }

\altaffiltext{2}{The Space Telescope Science Institute, 
3700 San Martin Dr., Baltimore, MD 21218.}

\altaffiltext{3}{Department of Physics and Astronomy, The Johns Hopkins University, 
3400 N. Charles St., Baltimore, MD 21218}

\altaffiltext{4}{California Institute of Technology, Mail Code 105$-$24, 
Pasadena, CA 91125}

\altaffiltext{5}{Department of Astronomy, Yale University, PO Box 208101, 
New Haven, CT 06520}

\altaffiltext{6}{Laboratory for Astronomy and Solar physics, Code 681, Goddard Space Flight Center, Greenbelt, MD 20771}

\setcounter{footnote}{6}
\begin{abstract}
We examine the evolution of the sizes and number densities of disk galaxies 
using the high resolution images obtained by the Great Observatories Origins 
Deep Survey (GOODS) with the Advanced Camera for Surveys (ACS) on the 
{\it Hubble Space Telescope (HST)}. The multiwavelength images are used 
to classify galaxies based on their rest--frame $B$-band morphologies
out to redshift $z\sim 1.25$. In order to minimize the effect of selection 
biases, we confine our analysis to galaxies which occupy the region of 
magnitude-size plane where the survey is $\sim$ 90\% complete at all 
redshifts. The observed size distribution is consistent with a log--normal 
distribution as seen for the disk galaxies in the local Universe and does 
not show any significant evolution over the redshift range 
$0.25\leq z \leq 1.25$. We find that the number densities of disk galaxies 
remains fairly constant over this redshift range, although a modest 
evolution by a factor of four may be possible within the 2$\sigma$ 
uncertainties. 
\end{abstract}


\section{Introduction}                   

Disk galaxies constitute about 60-80\% of the galaxies in the nearby 
Universe (Buta et al. 1994) and it is very important to understand how 
they formed and evolved. In the recent years, high resolution images from 
{\it HST} have proved extremely valuable in 
obtaining the structural parameters of galaxies out to $z\sim 1$. In this 
letter, we address the issue of whether the size distribution and number 
density of disk-dominated galaxies evolves with redshift. The evolution of 
disk galaxies has been explored previously via the magnitude--size 
($M_{B}-r_{e}$) relation (Schade et al. 1996; Lilly et al.1998; 
Roche et al. 1998; Simard et al. 1999); 
Bouwens \& Silk, 2002) and the Tully-Fisher ($M_{B}-V_{c}$) relation 
(Vogt et al. 1996). Based on {\it HST} imaging, most 
studies found evidence for a significant increase ($\sim$ 1--1.3 magnitude) 
in the rest--frame $B$-band surface brightness of disk galaxies
to $z=1$, while Simard et al. (1999) found no evidence for surface 
brightness evolution once the selection effects of the survey were taken 
into account. The luminosity--size evolution of disks remains
a controversial issue and the interpretation of any observed evolution
with redshift depends crucially on accounting for the selection biases of
the survey (Simard et al. 1999; Bouwens \& Silk, 2002). 
Lilly et al. (1998) reported that the abundance and size distribution 
remains constant out to $z \sim 1$, for the large disks with scalelengths 
greater than $\sim5$ kpc, for which their sample is fairly complete.
The space densities at different look-back times provides a key observable
to help determine how and when large galaxies like the Milky Way were formed.

The multi-wavelength ($B,V,i,z$) {\it HST}/ACS images from GOODS serve as 
an excellent resource to examine the size and number density evolution of 
disk galaxies with redshift. The availability of a long wavelength baseline 
from 4300\AA~ to 9000\AA~ allows galaxy properties to be studied consistently 
in the rest--frame $B$-band for galaxies out to $z\sim 1.25$. Also, 
the large area of the survey provides an ample number of galaxies over the 
range $0.25\leq z\leq 1.25$ for studying the number density 
evolution. We adopt the cosmology defined by $H_0$ = 70 \kms\ Mpc$^{-1}$, 
$\Omega_{M} = 0.3$, and $\Omega_{\lambda} = 0.7$ throughout this paper.

\section{Morphological Analysis and Identification of Disks}        

The analysis presented here is based on the first three 
epochs of observations of the Chandra Deep Field South (CDF-S) obtained 
via the GOODS program. The sample 
consists of all sources from the SExtractor (Bertin \& Arnouts 1996)
source catalogs of the GOODS CDF-S region (Giavalisco et al. 2004) 
with $z_{850} \leq$ 24.0 magnitude\footnote {$z_{850}$ denotes observed 
AB magnitude in the F850LP filter. The limiting isophote for the source 
detection was $\sim$ 27.3 magnitude/arcsec$^{-2}$ in the $z$-band.} and 
stellarity index $<$ 0.8. These criteria ensure that the signal-to-noise 
ratio is sufficient for the morphological analysis and excludes the stars. 
The photometric redshifts for the galaxies are from 
Mobasher et al. (2004) and are found to be robust 
($rms \leq 0.11$) out to $z\sim 1$ for objects with $z_{850} \leq$ 24.0
magnitude based on the comparison with available spectroscopic redshifts. 
The final sample consists of 2781 objects with $0.25\leq z\leq 1.25$.

We derive the structural parameters of the galaxies through a two-dimensional 
modelling of the surface brightness distribution using the 
GALFIT software (Peng et al. 2002). We use a single S\'{e}rsic (S\'{e}rsic 
1968) function to model the brightness profiles, and used simulations to 
verify that the S\'{e}rsic index, $n$, can provide a reliable 
classification of the bulge-dominated and disk-dominated galaxies even at 
the faint magnitudes. The quality of the best-fit model 
is judged based on the reduced chi square ($\chi^{2}_{\nu}$) value which 
should be close to unity when the model is a good match to the data. 
%
\vskip 0.3cm
\figurenum{1}
\psfig{file=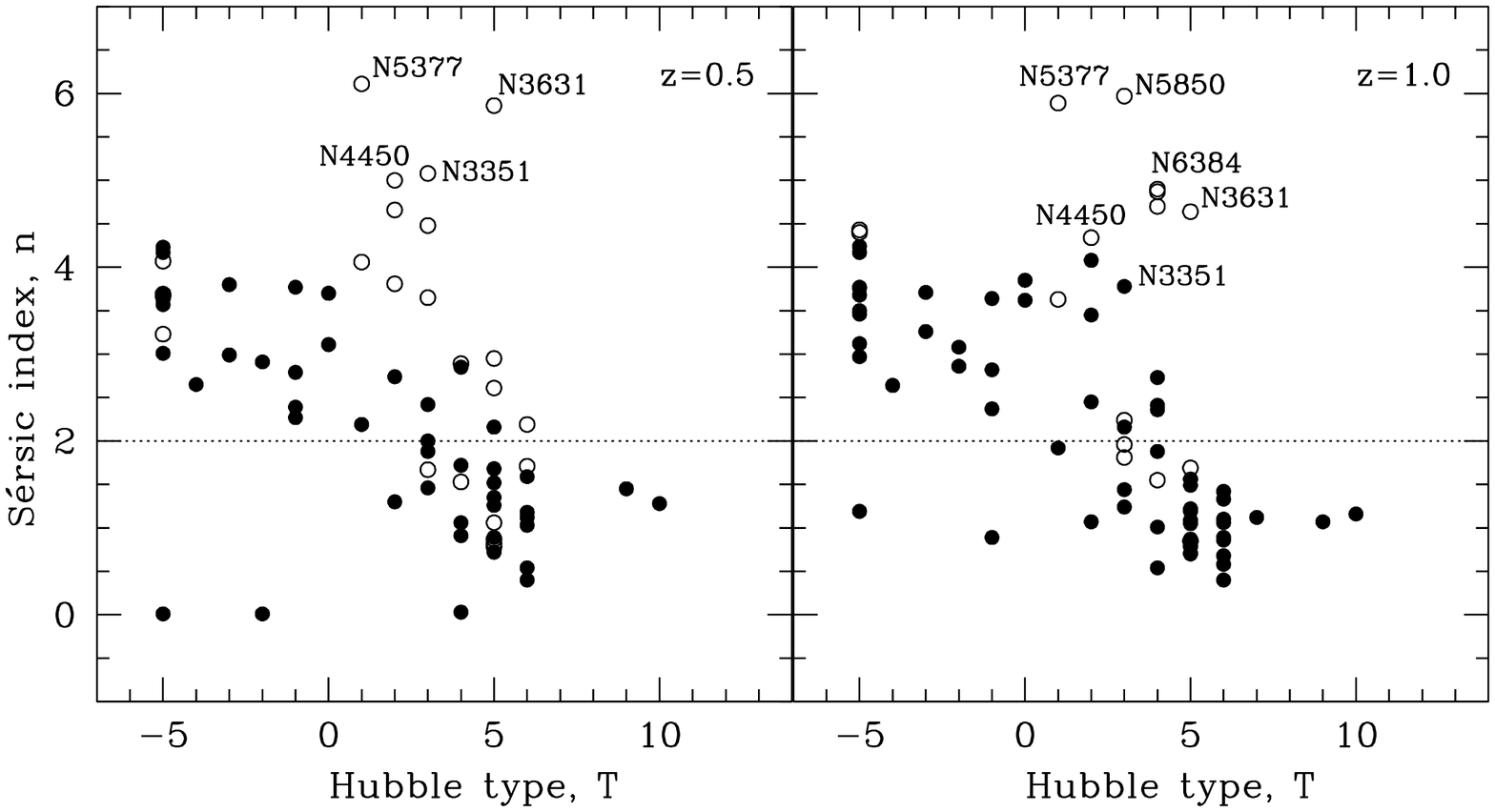,width=8.8cm,angle=90}
\figcaption[fig1.ps]{
The measured S\'{e}rsic index versus galaxy Hubble type from RC3
(de Vaucouleurs et al. 1991), for the redshifted Frei sample. Disk galaxies
are found to have $n < 2.0$ and bulges have $n > 2.0$. The open circles
represent galaxies which have low surface brightness, and constitute most
of the outliers in our classification scheme. The galaxies that deviate
most from their expected galaxy type are labelled, and in most cases have
a bright nucleus or star-forming regions which complicates the fits.
Objects with $n = 0$ have large $\chi^{2}_{\nu}$ values because the fits were
poor.
\label{fig1}}
\vskip 0.3cm
\noindent
The index, $n$, is known to correlate with galaxy morphology
(Andredakis, Peletier, \& Balcells 1995). 
In Figure 1, we illustrate the classification based on $n$, using 
nearby galaxies from the Frei sample (Frei et al. 1996), which 
have been redshifted artificially to $z=0.5$ and $z=1.0$ (Conselice 2003). 
The criterion $n < 2.0$ allows us to select disk-dominated galaxies (Sbc-Sdm)
even when they have morphological complexities such as dust, star-forming 
regions, etc. However, a few late-type galaxies have high $n$ due to 
the presence of a bright nucleus, or circumnuclear star formation in the 
center and may be missed by our disk selection criterion. It is 
encouraging that only few early-type galaxies migrate to the low $n$ values, 
and they usually have large $\chi^{2}_{\nu}$ values which implies poor fit
to the data.  

The structural parameters for the galaxies were measured 
in the rest-frame $B$-band at all redshifts. When the rest-frame $B$-band 
is redshifted to wavelengths that fall in the gap between two filters, we 
use an average of the measurements made in the two filters. We identified 
1508 disk galaxies, after excluding about 1\% of the galaxies with $n <2$ 
that have $\chi^{2}_{\nu} > 5$. Based on the spectral type used to assign 
photometric redshifts, we estimate about 3\% contamination from 
ellipticals and S0's within our sample of disk galaxies. We adopt 
the effective (or half-light) radius, $r_{e}$, from the S\'{e}rsic 
fit as a size measure. Errors on the measured sizes and magnitudes were 
estimated using extensive simulations done by introducing ``artificial'' 
galaxies with exponential disks and $r^{1/4}$ bulges into the GOODS images 
(Giavalisco et al. 2004). The simulated galaxies have bulge-to-disk 
($B/D$) ratios covering the range $-0.6 < $log$(B/D) < -1.8$ typical of 
Hubble types Sa to Sdm (Graham 2001), disk magnitudes 
$18 \leq z_{850} \leq 24$, disk half-light radii 
0\asec .1 $\leq r_{e} \leq $3\asec .0, and various ellipticities and 
position angles. We find that the systems with $n<2$ are predominantly 
galaxies with $B/D < 0.1$ and the bulge does not influence 
their derived disk parameters significantly. For galaxies with $n<2$, the 
mean difference in the integrated magnitudes and half-light radii between 
the recovered values using the S\'{e}rsic profile and the input disk 
parameters was $< \Delta z_{850} > = -0.14$ mag and 
$< \Delta $log$ r_{e} > = 0.06$ with dispersions of
$\sigma \Delta z_{850} = 0.40$ and $\sigma \Delta $log$ r_{e} = 0.08$. 
%
\vskip 0.3cm
\figurenum{2}
\psfig{file=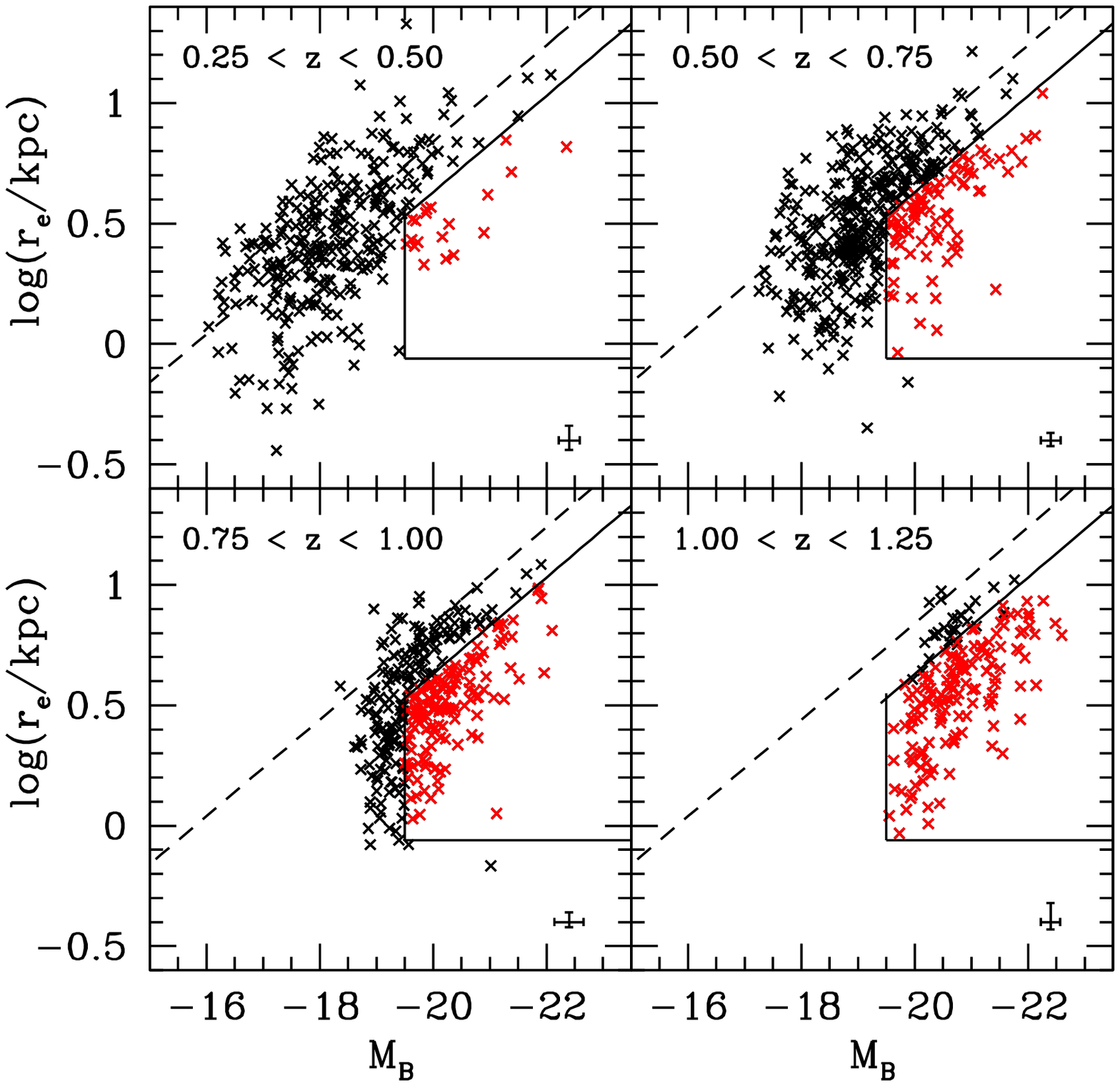,width=8.2cm,angle=0}
\figcaption[fig2.ps]{
The absolute $B$ magnitude versus half-light radius (M$_{B}-r_{e}$) relation
is shown for all the observed disk galaxies as a function of redshift. 
The trapezoidal region ({\it solid line}) is bounded by M$_{B} = -19.5$,
$r_{e} = 0.8$ kpc, and $\mu_{B}^{0}=20.6$ mag arcsec$^{-2}$, and encloses the
disk galaxies that are least affected by selection biases ({\it red crosses})
at all redshifts. In each panel, representative error bars are provided for a 
galaxy with median properties of the ``selection-free'' sample. The 
{\it dashed line} shows the Freeman relation for exponential disks with 
constant central surface brightness, $\mu_{B}^{0} = 21.65$ mag arcsec$^{-2}$.
\label{fig2}}

\section{Results}     

{\bf Magnitude$-$size relation for disk galaxies:}
In order to compare the evolution of disks at different redshifts it is 
important to consider the effect of selection biases which can artificially 
introduce evolutionary signatures in the absolute magnitude--size 
(M$_{B}-r_{e}$) plane shown in Figure 2. 
From our simulations we found that the selection in the $z$-band is more 
than 90\% complete for disk galaxies with surface brightness, 
$\mu_{z}^{e} \leq$ 23.4 mag arcsec$^{-2}$, measured within the effective 
radius. Assuming an exponential profile for the disks, this corresponds 
to a rest--frame $B$-band central surface brightness of 
$\mu_{B}^{0} \leq$ 20.6 mag arcsec$^{-2}$ in the highest redshift bin.
The $z_{850} \leq $ 24 magnitude criterion adopted in the sample selection 
translates to $M_{B} \leq -19.5$ in this redshift bin, and the smallest disk 
size is $r_{e} \sim 0.8$ kpc. Thus, disk galaxies 
with $M_{B} < -19.5$, $\mu_{B}^{0} \leq$ 20.6 mag arcsec$^{-2}$, 
and $r_{e} > 0.8$ kpc are almost free of selection biases at all redshifts 
and their distribution in the M$_{B}-r_{e}$ plane must reflect the actual 
luminosity or size evolution. Therefore, only disk galaxies satisfying the 
above criteria ({\it red points} in Figure 2) are considered for further 
analysis. At redshifts $z\geq 1$, the surface brightness threshold 
for 90\% completeness is about a magnitude brighter than the Freeman value. 
The small number of disks with luminosities $M_{B}< -21$ in the lowest 
redshift bin is striking even after accounting for the difference in 
comoving volumes in the various redshift bins. There is a class of high 
surface brightness disks with M$_{B} < -21.5$ and $r_{e} < 4$ kpc which 
become dominant at $z \geq 1$; a similar observation of high surface 
brightness disks at $z \geq 0.9$ was also reported by Simard et al. (1999). 
These objects could be among the most strongly evolving disk population 
from $z\sim 1$ to the present.

In Figure 3 the luminosity function (LF) is presented for the ``selection
-free'' disk galaxies in all redshift bins along with a representative 
``non-evolving'' LF\footnote {Although a direct comparison 
with LF of disk galaxies at $z=0$ (de Jong \& Lacey 2000) would be ideal, the 
difference in the sample selection makes such a comparison non-trivial and we 
defer such a comparison to a future paper. For simplicity, we adopt the same 
functional form for the model ``non-evolving'' LF to facilitate a comparison 
within the GOODS data set.}. Our selection criteria restricts the 
analysis to the most luminous disks and the observed LF at 
$z < 1 $ is well--represented by a non-evolving LF over the range 
of luminosities considered. However, at $z > 1$, the number of luminous 
disks with M$_{B} < -21$ is marginally higher than expected 
from the non-evolving LF and the observed points show a relative shift 
towards higher luminosities.
%
\vskip 0.3cm
\figurenum{3}
\psfig{file=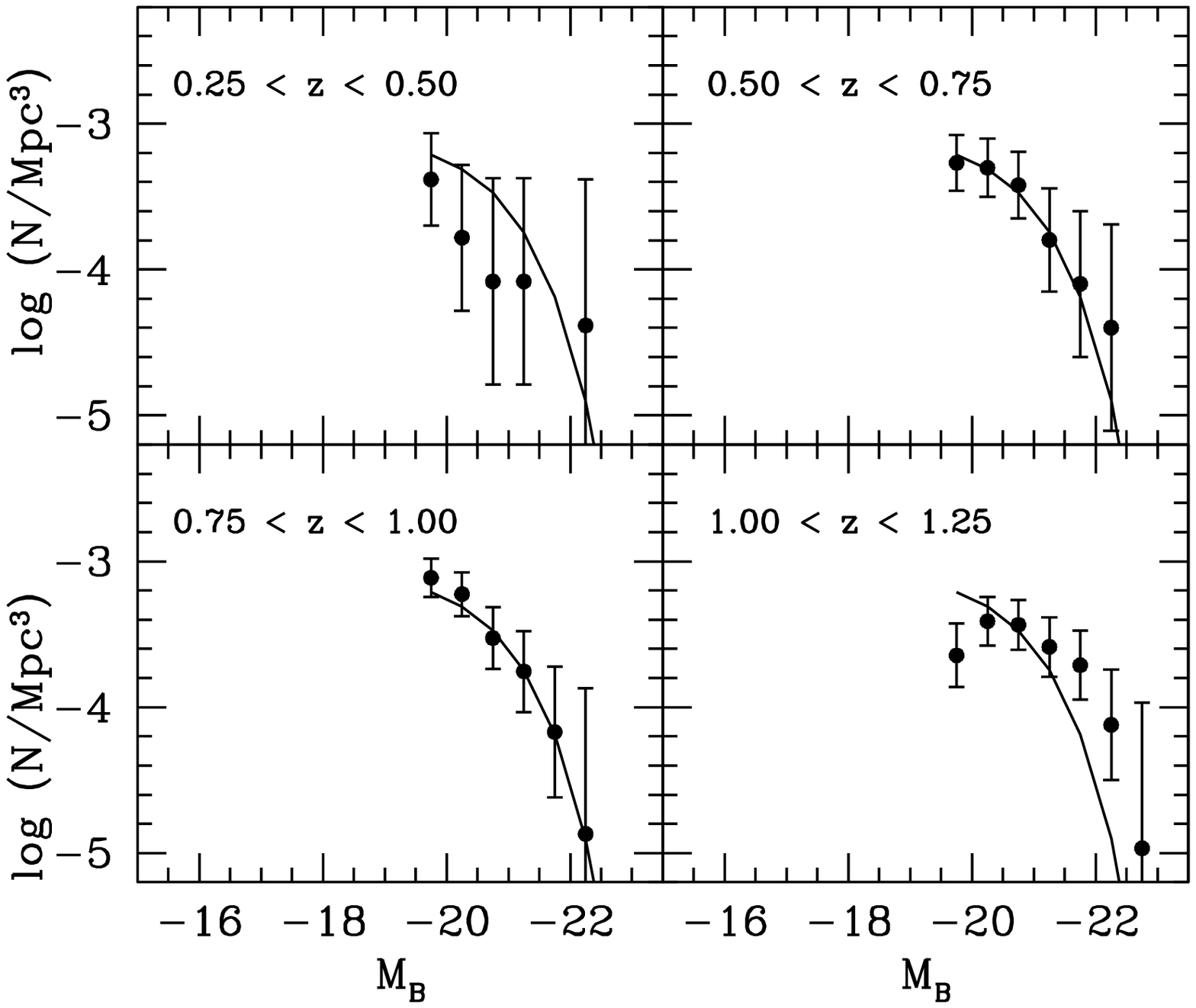,width=8.2cm,angle=0}
\figcaption[fig3.ps]{
The observed LF ({\it filled circles}) for disk galaxies
chosen to be {\bf free of selection biases} from figure 2, is presented along
with the Poisson error bars. A non-evolving LF
({\it solid line}) similar to that observed for local spiral disks
(de Jong \& lacey 2000) is also shown and is characterized by
M$_{B}^{*} = -20.6$, $\alpha = -0.90$. The normalization adopted for
the non-evolving LF has been adjusted to match the total
number of galaxies at $0.50 \leq z \leq 0.70$ and is held fixed for the
other redshift intervals to look for evolutionary signatures within the
sample.
\label{fig3}}
\vskip 0.3cm
\noindent

{\bf Size distribution of disk galaxies:}
Overall, the size distribution (SD) for disk galaxies is
consistent with a log-normal distribution and there is no noticeable size
evolution with redshift within our sample (Figure 4). To examine the effect of 
the selection and the measurement biases on the observed SD at various
redshifts within the sample, we carried out simulations by inserting 
``artificial'' disk galaxies in the GOODS images (see \S 2). 
The disk galaxies in the local Universe show a log--normal size distribution,
and luminosity-size relation, $r_{e} \propto L^{-\beta} \sim L^{1/3}$ 
(de Jong \& Lacey 2000). We adopt this analytical form and the LF discussed 
in the previous section, for the input non-evolving luminosity--size 
distribution function for galaxies in our simulations. 
After applying the same selection function to the simulated galaxies as done 
for observed galaxies, the sizes are re-measured for the ``selection free'' 
sample. At all redshifts, the surface brightness threshold adopted for the
selection affects the SD at large radii, and the peak of the distribution
shifts to smaller sizes. The measured sizes are in good agreement with the 
input sizes in the simulations for $z<1.00$, but are biased towards larger 
sizes at higher redshifts. For the non-evolving model this occurs because
galaxies become increasingly fainter at high redshifts and the uncertainties 
in the background estimation are larger.  
Thus, for the observed galaxies the SD is not significantly affected by the 
selection or measurement bias for small disk sizes ($r_{e} < 4$ kpc) at 
$z < 1$. But at higher redshifts, given the nature of the measurement bias 
an evolution in the SD due to the small disks cannot be distinguished from 
the effect of luminosity evolution which can make disks brighter and reduce 
the bias. 
\vskip 0.3cm
\figurenum{4}
\psfig{file=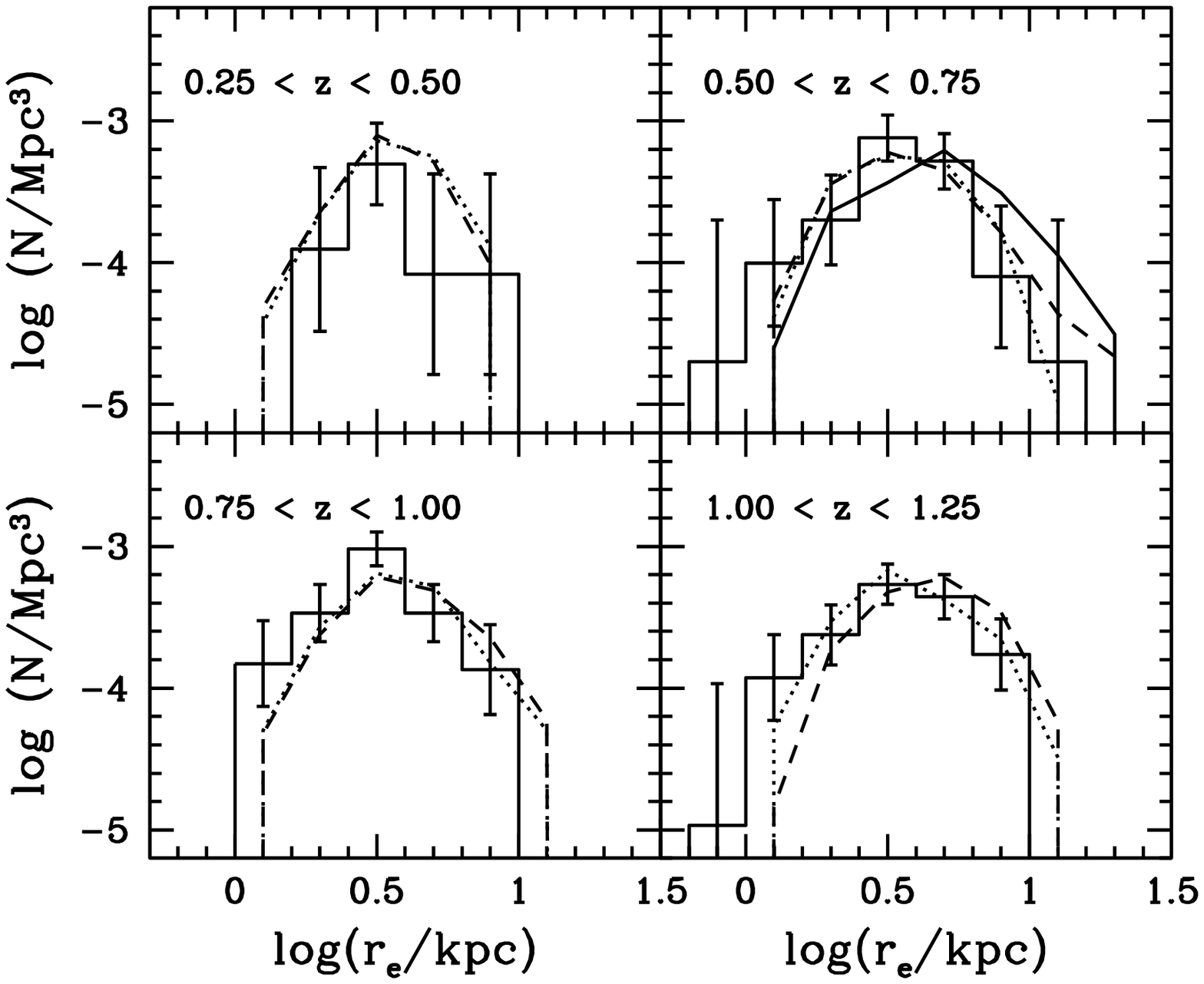,width=8.2cm,angle=0}
\figcaption[fig4.ps]{
The histogram shows the observed size distribution along with the Poisson
error bars for the disk galaxies which are {\bf free of selection biases}.
The curves show the effect of selection {\it dotted line} and measurement
{\it dashed line} on an input ``non-evolving'' SD {\it solid line} based
on simulations. The input log--normal distribution (prior to selection and
measurement) with a peak at $r_{e} = 6$ kpc and width $\sigma$(ln$r_{e}$)
$\sim$ 0.5 provides a good match to the size distribution observed at
$0.50 < z < 0.75$ after the selection criteria is applied, and is held
fixed for the other redshifts; the normalization has been adjusted
to match the number of galaxies in this redshift bin. The selection causes
the peak of the log-normal distribution to shift to smaller sizes
($r_{e} = 4$ kpc). Measurement biases are significant only in the highest
redshift bin.
\label{fig4}}
\vskip 0.3cm
\noindent

{\bf Number densities of disk galaxies:}
The observed number density of disk galaxies (Figure 5) that are free of 
selection biases in the GOODS CDF-S field, is found to remain almost 
unchanged out to redshift of $z=1.25$ within the uncertainties. We divided 
the sample into small disks ($r_{e} < 4$ kpc) and large disks 
($r_{e} \geq 4$ kpc) to look for differences in their relative abundances. 
Neither sample shows strong evolution within the uncertainties, except for 
a mild increase in the number of small disks at $z \sim 0.8$. Assuming that 
the evolution in number density can be expressed in the form, 
$n(z) \propto (1+z)^{\alpha}$, a formal linear regression analysis gives 
$\alpha$ = 0.08$\pm$0.23 for all disks, and $\alpha$ = 0.00$\pm$0.27 and 
0.20$\pm$0.39 for the small and large disks respectively. In all cases, the 
best fit favors an almost constant number density over the whole redshift 
range. However, a factor of four change in the number densities of small 
disks cannot be ruled out within the 2$\sigma$ uncertainties of the 
obtained fits. The same is true for the large disks if we exclude the lowest 
redshift bin where the uncertainty is large. Thus, ignoring the possible 
effects of luminosity evolution at $z>1$, it appears that the population of 
disk galaxies could have undergone only very modest evolution in their number 
densities from $z=0.25$ to $z=1.25$. 
In order to check for contamination from bulge-dominated galaxies that scatter 
into the sample at faint magnitudes we re-did the analysis using only those 
galaxies with $n \leq 1.5$, which have negligible bulge 
component and obtained very similar results.

%
\vskip 0.3cm
\figurenum{5}
\psfig{file=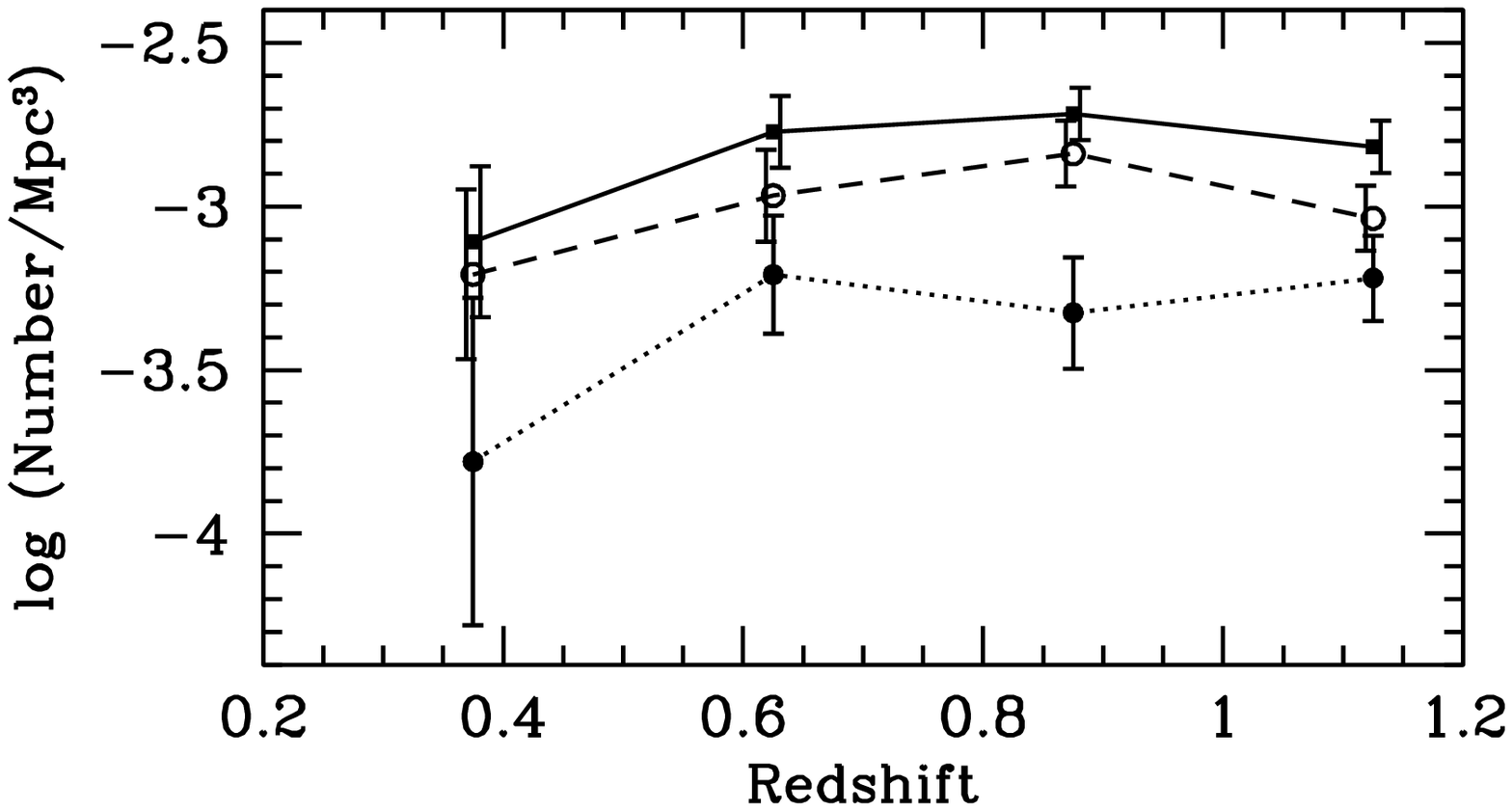,width=8.2cm,angle=0}
\figcaption[fig5.ps]{
The observed number densities of all disk galaxies ({\it filled squares,
solid line}) within the ``selection free'' region in Figure 2, is shown
as a function of redshift along with the statistical errors. The number
densities of small disks ($r_{e} < 4$ kpc, {\it open circles, dashed lines})
and large disks ($r_{e} \geq 4$ kpc, {\it filled circles, dotted lines})
do not show significant evolution in their relative abundance with redshift.
\label{fig5}}
\vskip 0.3cm
\noindent

\section {Discussion}          

In the context of evolution of disk galaxies in the M$_{B}-r_{e}$ plane, 
it is important to investigate the effects of $B$-band surface brightness
increase by $\sim$ 1.1$-$1.5 mag out to $z\sim 1$, 
claimed by some studies and contradicted by others (see \S 1). 
To allow direct comparisons at different redshifts within the GOODS data, 
we confined our analysis to galaxies with $M_{B} < -19.5$ and 
$\mu_{B}^{0} < 20.6$ mag arcsec$^{-2}$, such that they are not affected 
by selection biases in any redshift. 
For this sample, we do not see any significant evolution in the mean 
$B$-band surface brightness ($\Delta \mu_{B}^{0} < 0.4$). These results are 
in agreement with  Simard et al. (1999) who apply a uniform selection 
function from the highest to lowest redshift bins.
In contrast, the strong surface brightness evolution seen by 
Bouwens \& Silk (2002) is based on their comparison of samples that span 
different ranges of luminosity and surface brightness at different 
redshifts. If we include all the disks in our data within the 
apparent magnitude and surface-brightness limits for 90\% completeness in 
the $z$-band, we find an increase in the mean $B$-band surface brightness by 
$\sim 0.93$ mags from $z=0.2$ to $z=1$. Such a selection made based on 
apparent magnitude and size translates to different surface brightness 
thresholds in the intrinsic $M_{B}-r_{e}$ plane at different redshifts. 
For example, at low redshifts ($z < 0.5$) most of the disk galaxies chosen 
based on the above criteria have average $\mu_{B}^{0}$ close to the Freeman 
value, while at higher redshifts the average $\mu_{B}^{0}$ for the selected 
disks shifts to brighter magnitudes. 
The interpretation of the results in this case would thus depend on 
comparison to disk evolution models; alternatively a comparison 
can be made using like samples at different redshifts which is the approach 
used here.

The size distribution of disks over the redshifts $0.25 < z < 1.00$
is found to remain unchanged and exhibits a log--normal behavior similar 
to disk galaxies at low redshifts (de Jong \& Lacey 2000; Shen et al. 2003). 
This also agrees with the expectations from the hierarchical models of disk 
formation (Fall \& Efstathiou 1980). However, we note that the observed peak, 
and shape of the SD at large $r_{e}$ may be partly affected by the surface 
brightness threshold (Figure 2) that defines the selection.
The evolution of the observed SD at $z>1$ within the GOODS sample either due 
to a mild increase in the number of small-sized disks ($r_{e} < 4$ kpc) or 
luminosity evolution which makes the disks appear brighter at higher 
redshifts, cannot be distinguished given the measurement biases. 
Also, the scatter of bulge-dominated systems into the sample can lead to a 
similar effect and may pose a challenge to using the simple $n<2$ criteria 
for disk galaxy classification at high redshifts. 

From the observed number densities and luminosity-size
distributions, we conclude that the population of luminous 
(M$_{B} < -19.5$), disk galaxies with the sizes ranging from 0.8$-$10 kpc 
were present with roughly the same abundance at $z=1$ as at low
redshifts ($z \sim 0.2$) and are likely to have undergone only a modest 
luminosity evolution. This would require a formation epoch earlier than 
$z=1$ for these galaxies. Albeit based on a small number of galaxies, it is 
interesting that there is evidence for large disks at a mean redshift 
of $z = 2.3$ (Labb\'{e} et al. 2003) with roughly the same abundance as 
seen for the large disks at $z\sim 1$ in our data. 
Whether this implies that massive systems like the 
Milky Way galaxy were already in place at that epoch can be addressed with 
additional kinematic information to constrain the 
mass--to--light ratios. 
A more detailed analysis based on the full GOODS data will be presented in a 
future paper.

\acknowledgements
This work was supported by grant GO09583.01-96A from the Space Telescope 
Science Institute, which is operated by AURA under NASA contract NAS5-26555. 
We thank C. Y. Peng and T. Dahlen for useful discussions, and the referee
for the helpful comments.

\end{document}